\documentstyle[prl,aps,psfig,multicol]{revtex}
\begin{document}
\draft
\title{Magnetic Field Suppression of the Conducting Phase in Two Dimensions}
\author{D.~Simonian, S.~V.~Kravchenko, and M.~P.~Sarachik}
\address{City College of the City University of New York, New York, NY 10031}
\date{\today}
\maketitle
\begin{abstract}
The anomalous conducting phase that has been shown to exist in zero 
field in dilute two-dimensional electron systems in silicon MOSFETs 
is driven into a strongly insulating state by a magnetic field of about 
20~kOe applied parallel to the plane.  The data suggest that in the limit of 
$T\rightarrow0$ the conducting phase is suppressed by an arbitrarily 
weak magnetic field.  We call attention to striking similarities to magnetic 
field-induced superconductor-insulator transitions.
\end{abstract}
\begin{multicols}{2}

Recent experiments in high-mobility Si MOSFETs have provided strong evidence 
that a conducting phase exists in dilute two-dimensional (2D) electron systems 
in the absence of a magnetic field, in disagreement with predictions of the 
scaling theory \cite{g4} for non-interacting electrons.  We attribute this 
finding to the availability of samples of unusually high mobility, allowing a 
transition from insulating to conducting behavior with increasing 
electron density, $n_s$, at small densities 
($n_s\sim10^{11}\text{cm}^{-2}$).  We note that since the Fermi 
energy $\epsilon_F\propto n_s$ in two dimensions and the electron 
correlation energy $\epsilon_{ee}\propto n_s^{1/2}$, the ratio 
$\epsilon_{ee}/\epsilon_F$ is proportional to $n_s^{-1/2}$; therefore, 
the lower the electron density, the greater the role of 
electron-electron interactions.  For the 2D electron system in 
silicon, it has been shown experimentally \cite{cgap} that 
the temperature ($T$) and electric field ($E$) dependences of the 
resistivity on the 
far-insulating side of the transition are consistent with the 
presence of a Coulomb gap in the density of states, 
indicating that electron correlations play a significant role.  
Moreover, comparison of temperature scaling and electric field 
scaling \cite{prl96} near the $H=0$ transition in Si MOSFETs 
yields a dynamical exponent, $z\approx0.8$, close to the value $z=1$ 
expected theoretically for a strongly interacting system 
(see, {\it e.g.}, Ref.\cite{girvin}), again pointing to the importance of 
Coulomb interactions.  Strong electron-electron interactions 
may thus be a central feature that allows the existence of a 
conducting phase in two dimensions.  However, the nature of this 
phase remains unclear.

The influence of a magnetic field applied 
perpendicular to the plane of the 2D electron system has 
been studied in detail by Pudalov and coworkers 
\cite{pudalov} in high-mobility MOSFETs with comparable 
electron densities.  In these studies, the magnetoconductance 
is largely dominated by orbital effects which lead to the 
quantum Hall effect.  In this Letter we report the results of 
measurements of the resistivity in a magnetic field applied 
parallel to the plane; here the magnetic field couples 
to the spins, but not to the orbital motion.  Our results 
indicate that a parallel magnetic field has a dramatic effect 
on the transition, entirely eliminating the conduction mechanism responsible 
for the existence of the $H=0$ conducting phase above 
$\sim$~20~kOe.  Based on our data, we suggest that the 
conducting phase is suppressed by an arbitrarily weak 
magnetic field in the limit $T\rightarrow0$.  We point out 
further that the behavior in a magnetic field, as well as 
the critical behavior in zero field 
\cite{prl96,krav}, bear a strong resemblance to 
behavior reported near the superconductor-insulator 
transition in thin metal films \cite{hebard,goldman,kapitulnik}, 
raising the possibility that the anomalous conducting phase 
found in the 2D electron system in silicon MOSFETs is, in 
fact, a superconducting phase.

We report results of measurements of the linear and nonlinear DC resistivities 
of a high-mobility ($\mu^{max}_{T=4.2K}\approx24,000$~cm$^2$/Vs) Si 
MOSFET sample; data for two other high-mobility samples 
gave similar results.  We note that equivalent information is 
obtained from the temperature-dependence of the linear resistivity 
(in the limit $E\rightarrow0$) and the electric-field-dependence of the 
(nonlinear) resistivity in the limit $T\rightarrow0$, as was demonstrated 
\cite{prl96} by similar behavior near the critical point in the two cases.  
Measurements as a function of electric field are easier to perform and entail smaller errors.  As in earlier experiments, the electron density was 
set by adjusting the gate voltage.  The resistivity was 
measured as a function of parallel magnetic field, at various 
temperatures, and for different values of the electric field 
(determined by the measuring current).  No difference was 
found for in-plane magnetic fields applied parallel and 
perpendicular to the measuring current.  The samples and 
measurements are described in more detail in 
Refs.\cite{prl96,krav}.

Figure \ref{1} shows the nonlinear resistivity in units of $h/e^2$ 
as a function of electric field in a magnetic field of 5 kOe at a 
temperature of 0.1~K.  Each curve corresponds to a 
different electron density (gate voltage).  For comparison, the inset shows 
the resistivity as a function of electric field in the absence of 
a magnetic field for comparable electron densities.  In zero magnetic 
field, the curves clearly separate into two groups: for low electron 
densities the resistivity increases with decreasing 
temperature (insulating behavior), while for higher electron 
densities the resistivity decreases with decreasing 
temperature (conducting behavior); the resistivity at the 
transition ($n_s=n_c$) is independent of electric field and 
approximately equal
\vbox{
\vspace{0.15in}
\hbox{
\hspace{.4in}
\psfig{file=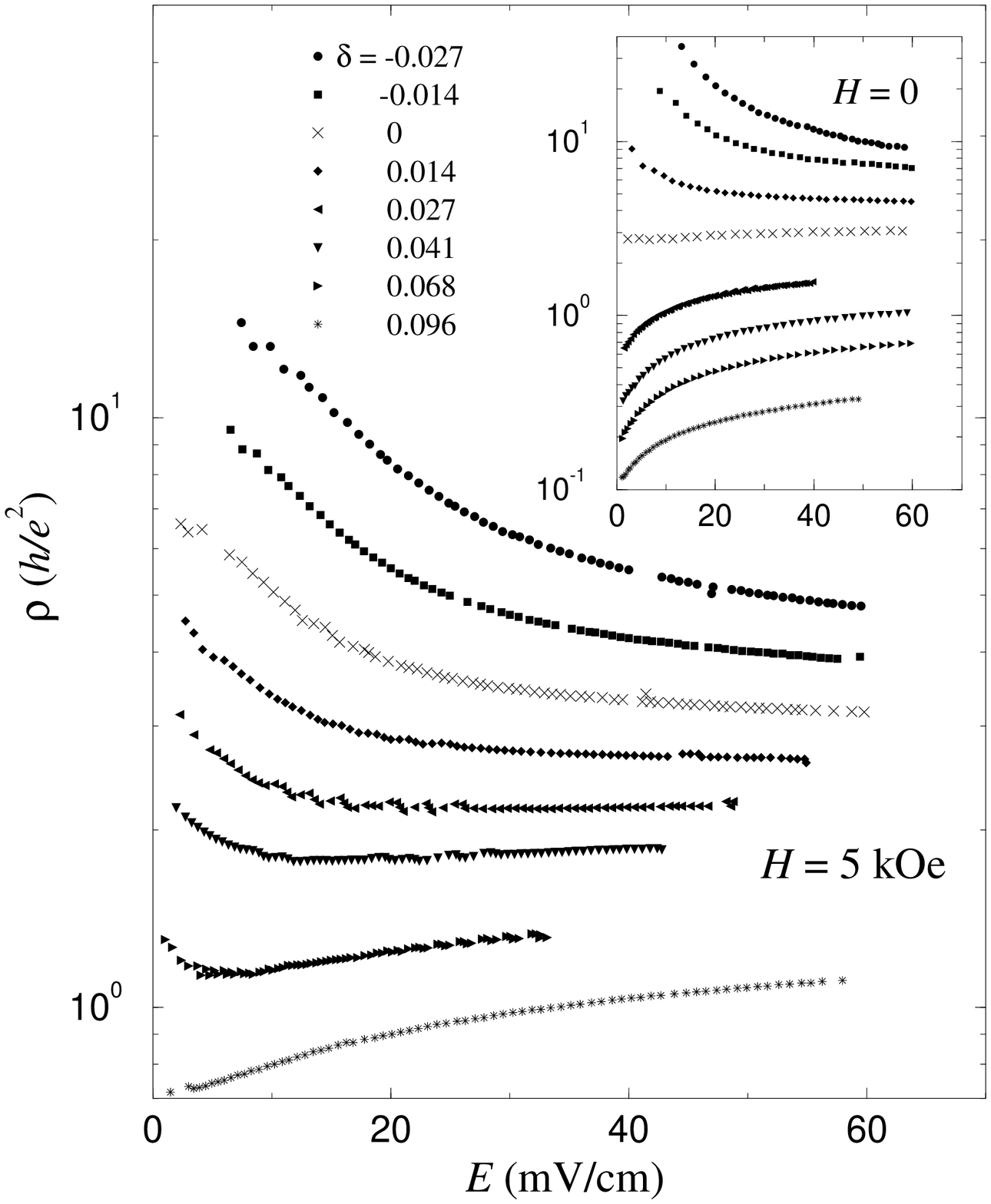,width=3.1in,bbllx=1.5in,bblly=1in,bburx=7.75in,bbury=9.25in,angle=0}
}
\vspace{0.15in}
\hbox{
\hspace{-0.15in}
\refstepcounter{figure}
\parbox[b]{3.4in}{\baselineskip=12pt \egtrm FIG.~\thefigure.
Resistivity as a function of electric field on a semilogarithmic scale at $H_{||}$ = 5 kOe and $T=0.10$~K.  Electron densities are specified relative 
to the $H=0$ critical density, $n_c=8.03\times10^{10}$~cm$^{-2}$; 
$\delta\equiv(n_s-n_c)/n_c$.  The inset shows $\rho(E)$ in the absence of a 
magnetic field at $T=0.22$~K, for $\delta=-0.065, -0.050, -0.030, 0, 0.052, 
0.10, 0.16,$ and $0.27$.  The crosses correspond to $\delta=0$.
\vspace{0.10in}
}
\label{1}
}
}
to $3h/e^2$.  As demonstrated in reference 
\cite{prl96}, a single (horizontal) multiplicative factor can be used to 
obtain scaling.  The effect of a parallel magnetic field is 
clearly shown in the main part of Fig.~\ref{1}: a magnetic field of 
5~kOe drives all curves 
toward more insulating behavior.  Moreover, there is a 
qualitative change: for some electron densities the 
resistivity exhibits non-monotonic behavior, developing a 
shallow minimum.  We shall return to this point below.

The resistivity is shown on a logarithmic scale as a function of magnetic 
field at a fixed temperature of 0.25~K in Fig.~\ref{2} for three 
different electron densities on the conducting 
side of the $H=0$ transition ($n_s>n_c$).  The resistivity initially stays 
approximately constant up to $H_{||}\approx4$~kOe;
data at low fields are shown on an expanded scale in
the inset to Fig.~\ref{1} for an electron density corresponding to 
$\delta\equiv(n_s-n_c)/n_c=0.15$.  
The resistivity then increases sharply as the magnetic field is 
raised further, changing by almost three orders of 
magnitude.  Above $H_{||}\sim20$~kOe, it saturates and stays 
approximately constant up to the highest measured field, $H_{||}=70$~kOe. 
A parallel magnetic field has dramatically altered 
the
\vbox{
\vspace{.035in}
\hbox{
\psfig{file=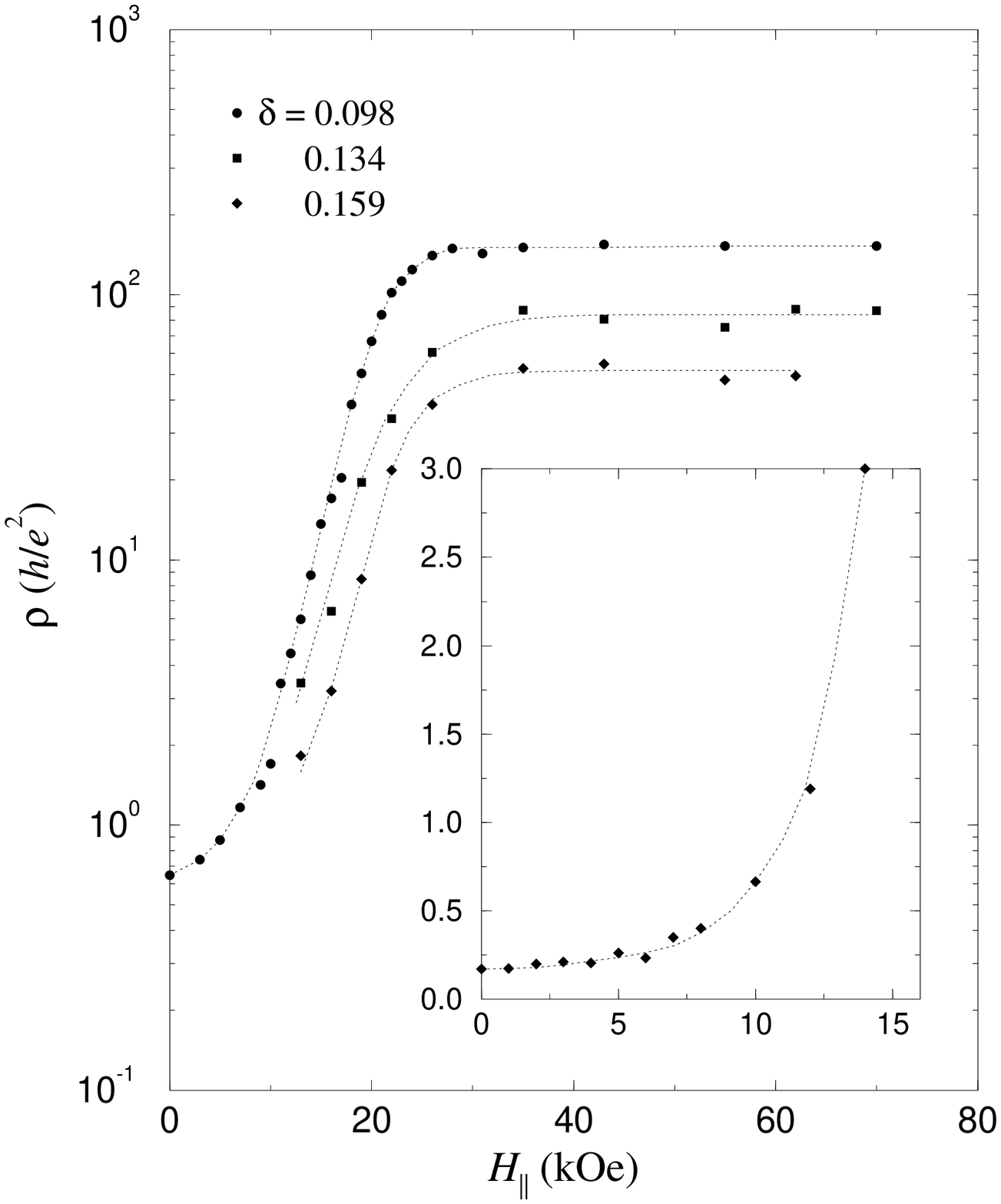,width=3.3in,bbllx=.5in,bblly=1.25in,bburx=7.25in,bbury=9.5in,angle=0}
}
\vspace{0.3in}
\hbox{
\hspace{-0.15in}
\refstepcounter{figure}
\parbox[b]{3.4in}{\baselineskip=12pt \egtrm FIG.~\thefigure.
Resistivity on a logarithmic scale as a function of a magnetic field applied 
parallel to the plane at $T=0.25$~K for three electron densities.  The inset 
shows the resistivity on a linear scale in small magnetic field for 
$\delta=0.15$ and $T=0.13$~K.
\vspace{0.10in}
}
\label{2}
}
}
system, apparently suppressing the conduction mechanism in the anomalous 
conducting phase entirely in fields above 20~kOe.  The behavior is strongly 
reminiscent of the quenching of superconductivity by a magnetic field (except, 
of course, that the zero-field resistivity in our case is finite rather 
than zero). The Zeeman energy, $g\mu_B H_{||}$, at 20~kOe corresponds to a 
thermal energy $k_BT_H$ with $T_H=2.7$~K. Note that 
$T_H\sim T^*\approx2$~K where $T^*$ marks the onset of the low-temperature 
conducting phase in zero field (see the lowest curve of Fig.~\ref{4}).
 
Measurements in magnetic fields oriented perpendicular to 
the plane of the electrons confirm earlier detailed magnetotransport 
results obtained  by Pudalov {\it et al}\cite{pudalov} 
in Si MOSFETs with comparable 
electron densities and mobilities.  Fig.~\ref{1} of their paper 
shows that the 
resistance is essentially constant up to 5~kOe, above which it 
rises sharply before it is overwhelmed by the quantum Hall 
effect above $\sim15$~kOe.  This puzzling, sharp initial increase has 
been the subject of some debate.  We 
suggest that its origin is the same as for a parallel field: a 
conducting phase exists at low temperatures which is 
suppressed by a magnetic field.  Thus, the anomalous $H=0$ conducting 
state is driven into a strongly insulating (``normal'') 
state either by $H_{||}$ or by $H_\perp$, in a qualitatively similar way.

We now consider whether one can identify a critical
\vbox{
\vspace{-0.25in}
\hbox{
\hspace{0.10in}
\psfig{file=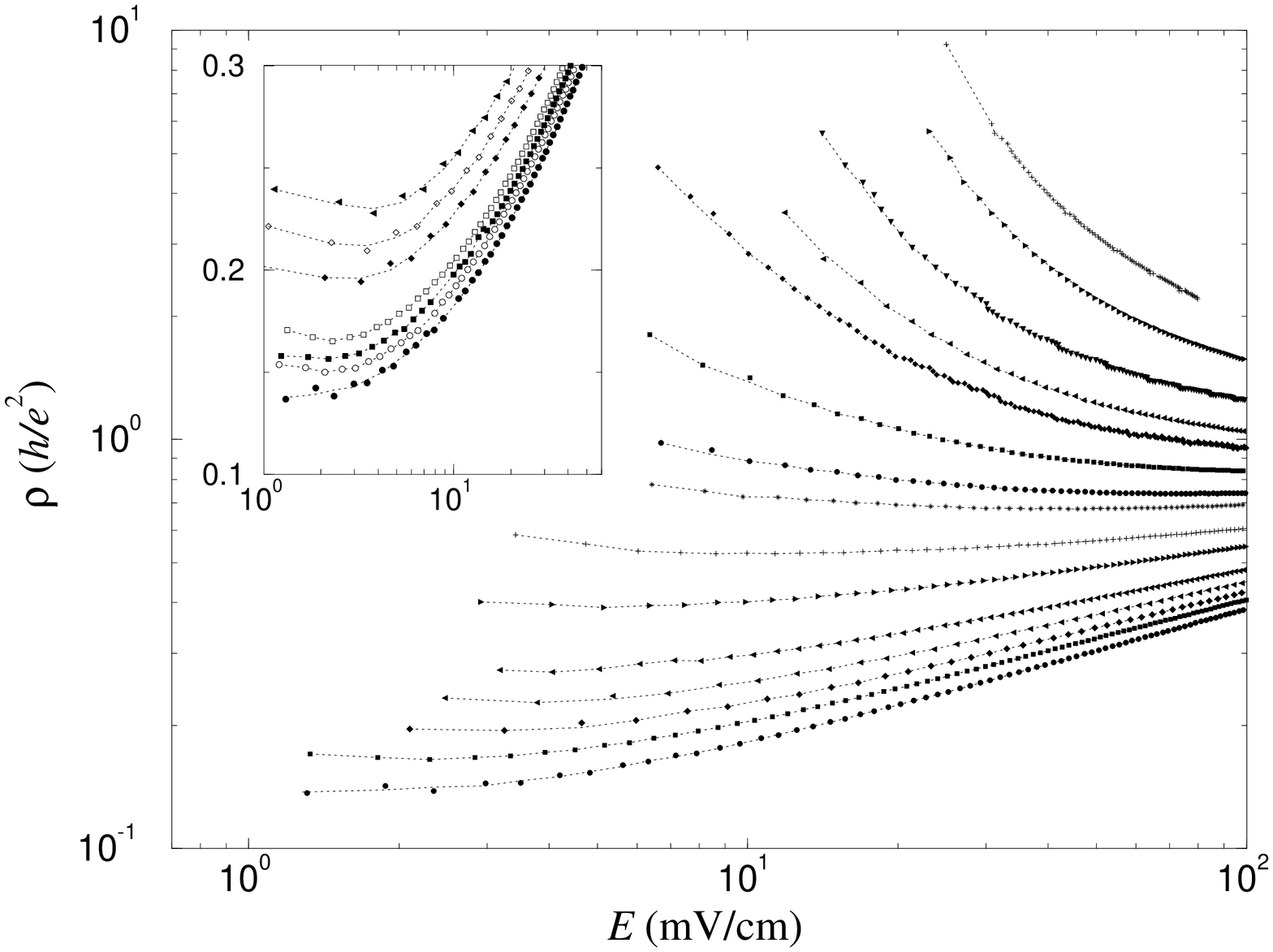,width=3.3in,bbllx=.5in,bblly=1.25in,bburx=7.25in,bbury=9.5in,angle=-90}
}
\vspace{0.3in}
\hbox{
\hspace{-0.15in}
\refstepcounter{figure}
\parbox[b]{3.4in}{\baselineskip=12pt \egtrm FIG.~\thefigure.
Isomagnetic curves of nonlinear resistivity as a function of electric field 
on a log-log scale for fixed electron density, $\delta=0.3$, at $T=0.10$~K. 
Each curve corresponds to a different value of parallel magnetic field, 
$H_{||}=0$ (bottom curve), $6, 8, 10, 12,$ 
$15, 17, 19, 20, 22, 24, 25, 27, 30,$ and $34$~kOe.  Minima in the 
resistivity are clearly illustrated in the inset, where data are shown on a 
linear scale for $H_{||}=0$ (bottom curve), $4, 5, 6, 8, 9,$ and $10$~kOe.
\vspace{0.10in}
}
\label{3}
}
}
parallel magnetic field below which the system is a 
conductor, and above which it is an insulator.  In Fig.~\ref{3}, we 
plot the nonlinear resistivity, $\rho(E)$, for a fixed electron 
density (corresponding to a zero-field
$\delta=0.3$) at 0.1~K.  Here each curve corresponds to a 
different value of $H_{||}$. As noted above, the curves are 
{\it qualitatively} different from those in zero field shown in the 
inset to Fig.~\ref{1}: the curves for $\delta>0$ display a shallow 
minimum in finite magnetic field, and it is no longer possible 
to use a single parameter to collapse them onto two separate branches, 
insulating and conducting, as was done at $H=0$\cite{prl96,krav}. 
Moreover, there is no universal ``critical'' value of the resistivity, $\rho(H_{||c}$).  This suggests that any finite magnetic field (at $T=0$)  drives the system into the insulating phase.

Finally, Fig.~\ref{4} shows the linear resistivity (at $E\rightarrow0$) as a 
function of temperature for a fixed electron 
density on the conducting side of the $H=0$ transition ($\delta=0.1$) in 
several parallel magnetic fields between 0 and 14~kOe.  The 
zero-field curve is typical of a conductor, with resistance 
dropping sharply as the temperature is decreased below 
$\approx2$~K, while at $H=14$~kOe it is strongly insulating.  Note that 
the magnetic field has almost no effect on the resistivity 
above $T^*\approx2$~K, while below $T^*$ the effect of $H_{||}^*$ is 
enormous (as discussed earlier, $T^*$ is the characteristic 
temperature below which the conducting phase exists in zero 
field).  We note the presence of resistivity 
minima at intermediate magnetic fields.  Again, one-parameter scaling with 
temperature breaks down, as did 
one-parameter scaling with electric field (see above).

The effect of a parallel magnetic field in the 2D
\vbox{
\vspace{-0.044in}
\hbox{
\psfig{file=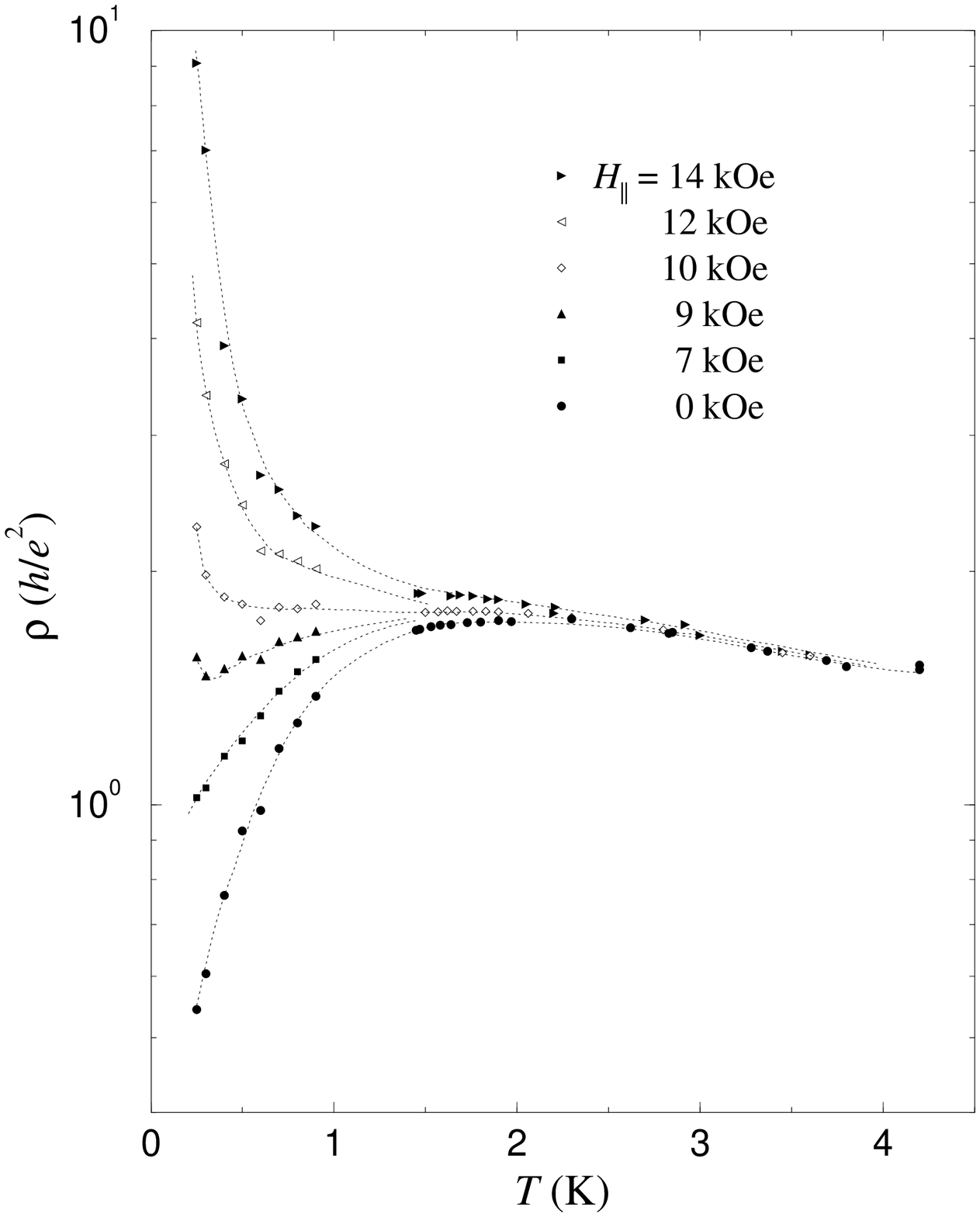,width=3.3in,bbllx=.5in,bblly=1.25in,bburx=7.25in,bbury=9.5in,angle=0}
}
\vspace{0.15in}
\hbox{
\hspace{-0.15in}
\refstepcounter{figure}
\parbox[b]{3.4in}{\baselineskip=12pt \egtrm FIG.~\thefigure.
Linear resistivity versus temperature in five different parallel
magnetic fields. The electron density corresponds to $\delta=0.10$.
\vspace{0.10in}
}
\label{4}
}
}
electron gas in Si MOSFETs 
strongly resembles the effect of a field in systems that undergo magnetic 
field-induced superconductor-to-insulator transitions (SIT).  For example, 
the isomagnetic curves of Fig.~\ref{4} are similar to those in Fig.~\ref{1} of 
Ref.\cite{hebard} measured near an SIT driven by a magnetic field in 
disordered indium oxide films. They are similar also to the temperature 
dependence of the resistivity of a bulk high-$T_c$ compound\cite{boebing}.

The possibility of superconductivity in Si MOSFETs has been considered by 
Takada\cite{takada} and by Hanke and Kelly\cite{hanke}.  More recently, 
$p$-wave superconductivity in this system has been proposed by Phillips and 
Wan\cite{phillips}.  In addition, various kinds of instabilities in 2D have 
been proposed theoretically (for review, see Ref.\cite{ando}), including 
Wigner crystallization, a transition to a ferromagnetic state at low electron 
densities, single-valley occupancy, and instabilities toward a charge-density 
or spin-density ground state.

To summarize, we report that a parallel magnetic field suppresses 
the anomalous conducting phase found at $H=0$ in the 2D electron 
system in Si MOSFETs.  The resistivity increases by several 
orders of magnitude at low temperatures, saturating above $\approx20$~kOe.  
Qualitatively similar behavior is found\cite{pudalov} in perpendicular field, 
which couples to orbital motion as well as spin, up to 
approximately 15~kOe; at higher perpendicular fields the magnetoconductance 
is overwhelmed by the quantum Hall effect.  The fact that a parallel magnetic 
field has such a dramatic effect indicates that the electrons' spins play a 
central role.  The fact that the Zeeman energy $g\mu_B H$ and thermal 
energy $k_BT$ that destroy the conducting phase are roughly comparable 
further supports this possibility.  One-parameter scaling 
with temperature and electric field, found to hold when $H=0$, breaks down 
even in a weak magnetic field, suggesting the elimination of the conducting 
phase by an arbitrarily small $H$.  The magnetoresistance strongly resembles 
that observed at the superconductor-to-insulator transition driven by a 
magnetic field.

We thank R.~N.~Bhatt, V.~Dobrosavljevi\'{c}, A.~B.~Fow\-ler, 
P.~Phillips, S.~Sachdev, and S.~L.~Sondhi for valuable discussions and 
Sera Cremonini for help in the analysis of the experimental data.  
This work was supported by the US Department of Energy under Grant 
No.~DE-FG02-84ER45153.

\end{multicols}
\end{document}